\documentclass{PoS}
\usepackage{amsmath}

\title{Charged and neutral Higgs bosons in final states with six bottom quarks}

\ShortTitle{Charged and neutral Higgs bosons in final states with six bottom quarks}

\author{\speaker{Radovan Dermisek}\\
        Physics Department, Indiana University\\
        E-mail: \email{dermisek@indiana.edu}}

\author{Enrico Lunghi\\
        Physics Department, Indiana University\\
        E-mail: \email{elunghi@indiana.edu}}
        
\author{Navin McGinnis\\
        High Energy Physics Division, Argonne National Laboratory\\
        E-mail: \email{nmcginnis@anl.gov}}
        
\author{Seodong Shin\\
        Department of Physics, Jeonbuk National University\\
        E-mail: \email{sshin@jbnu.ac.kr}}        

\abstract{In extensions of two Higgs doublet models with vectorlike quarks, the decays of vectorlike quarks may be easily dominated by cascade decays through charged or neutral Higgs bosons leading to signatures with 6 top or bottom quarks. Since top quark decays also contain bottom quarks, the 6 bottom quarks in final states is a common signature to a large class of possible decay chains. We present a search strategy focusing on this final state and find the mass ranges of vectorlike quarks and Higgs bosons that can be explored at the Large Hadron Collider. Among other results the sensitivity to the charged Higgs boson, extending above 2 TeV, stands out when compared to models without vectorlike matter.}

\FullConference{ICHEP 2020\\
                July 28 - August 6, 2020\\
                Prague}

\begin{document}

\section{Introduction}
In the search for physics beyond the Standard Model, model building strategies span a wide range of goals from purely phenomenological explorations to probing the UV dynamics in nature. Among the conceptually simplest of these extensions, extra multiplets of vector-like fermions and heavy Higgses are a common feature. Experimental probes of these particles typically focus on decay modes of new particles involving only SM states. For examples, see~\cite{Aad:2019zwb,Sirunyan:2019arl,Aaboud:2018xuw,Aaboud:2018saj}. In general, models with extended Higgs \textit{and} matter sectors will also involve their mixing with SM particles leading to additional decay modes.

% Including the so-called cascade decay modes can largely alter the interpretation of existing experimental limits while offering new opportunities for search strategies to be explored. In these proceedings, we discuss the advantages of cascade decay channels in a 2HDM with vector-like quarks that can mix with the third generation quarks in the SM and report on the possible reach of the masses of both vector-like quarks and heavy Higgses at the High Luminosity LHC.
 
 In these proceedings, we discuss the advantages of cascade decays of vector-like quarks through heavy Higgses in a 2HDM where the third generation quarks in the SM can mix with vector-like quark doublets and singlets. Including these cascade decay modes can largely alter the interpretation of existing experimental limits while offering new opportunities for search strategies to be explored. We report on the possible reach of the High Luminosity LHC for masses of both vector-like quarks and heavy Higgses.

As an example of a UV completion of the SM with an extended Higgs sector and vector-like quarks, we briefly discuss the minimal Supersymmetric Standard Model (MSSM) extended with a complete vector-like family. In this model, the seven largest couplings of the SM can be simply understood from their IR fixed points in the renormalization group flow of this model, considering only a single scale of new physics.

\section{Cascade decays of vector-like quarks through heavy Higgses}

A type-II 2HDM with vector-like $SU(2)$ doublet and singlet quarks with general mixing to third generation quarks was considered in~\cite{Dermisek:2019vkc}. The resulting decay modes of new top- and bottom-like quarks include decays to heavy neutral Higgses, e.g. $t_{4}\rightarrow H t$ or $b_{4}\rightarrow H b$, and charged Higgses, $t_{4}\rightarrow H^{\pm} b$ or $b_{4}\rightarrow H^{\pm} t$, in addition to typical decay modes involving $Z, W$ or $h$ bosons. Small mixing with SM quarks can lead to typical scenarios where heavy Higgs decays dominate over $Z, W$ or $h$ decays in large regions of parameters. For more details of the model, couplings and decay modes discussed here see~\cite{Dermisek:2019vkc}.

\begin{figure}[t]
\begin{minipage}{6in}
\centering
	\raisebox{-0.5\height}{ \includegraphics[scale=0.5]{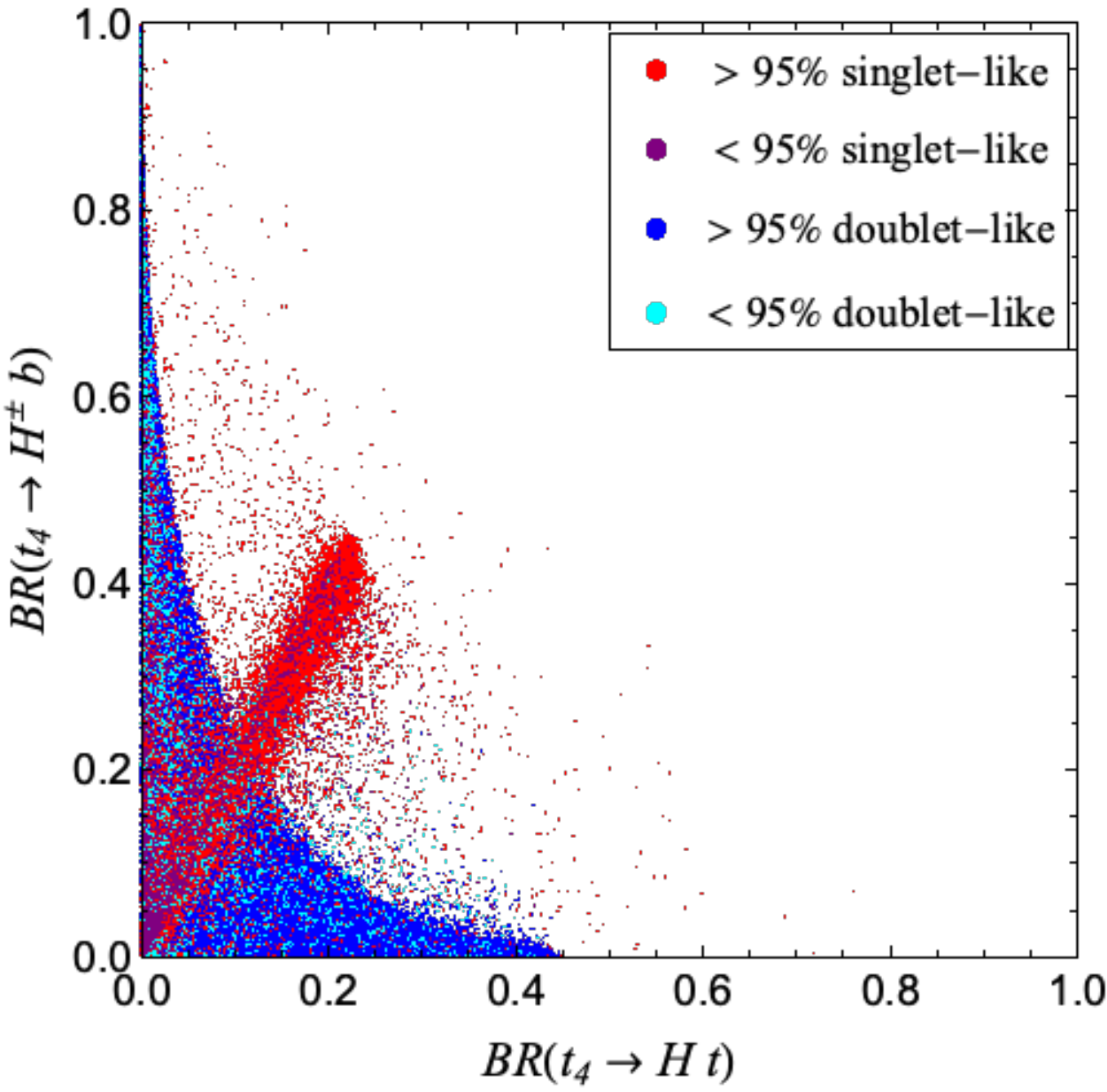}
	 \includegraphics[scale=0.5]{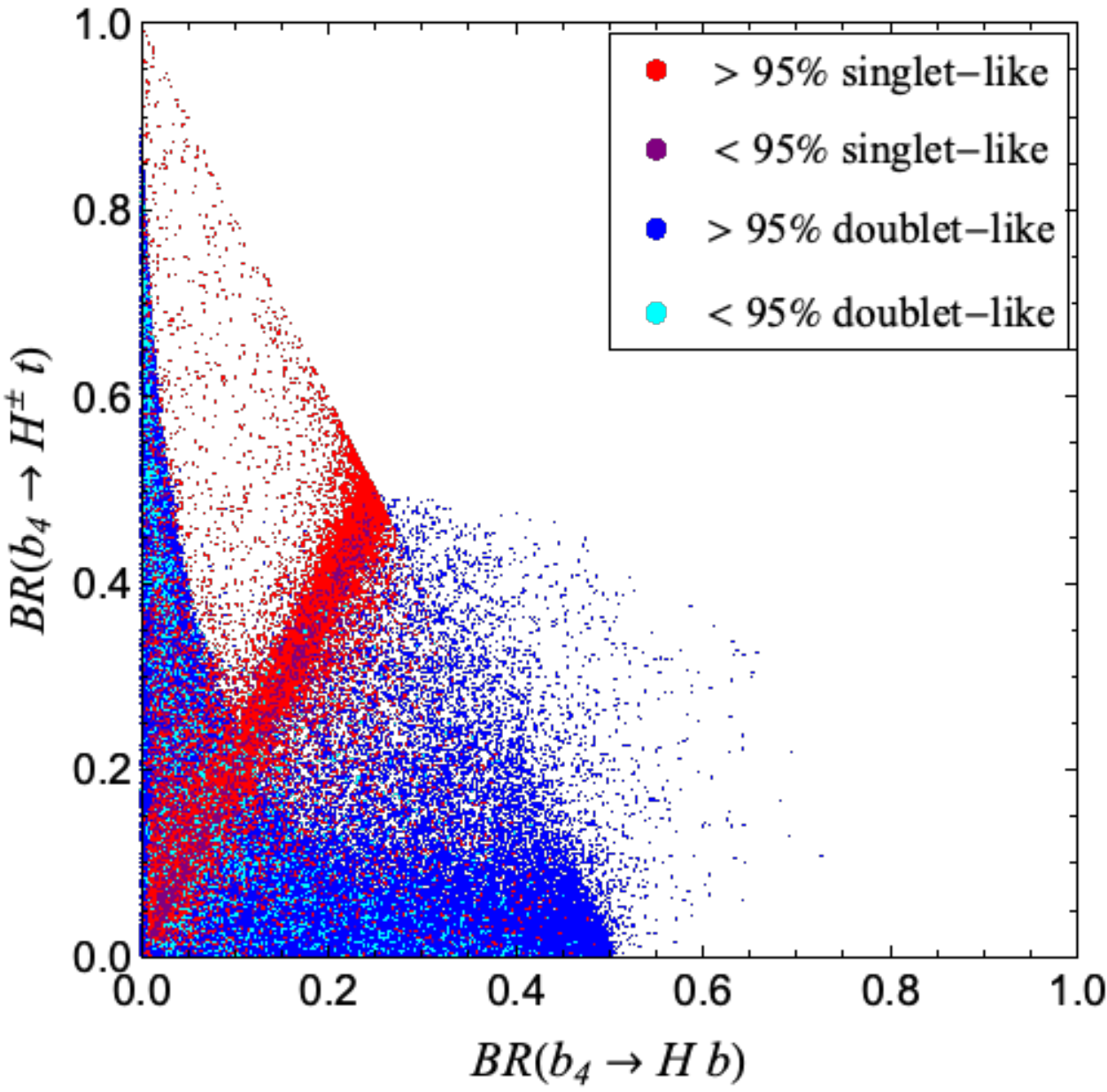}}	
\caption{$t_4$ and $b_4$ branching ratios into heavy charged and neutral Higgses in the general scenario where mixing of vector-like quarks to third generation is allowed. Mixing to new doublet- and singlet-like quark mulitplets is considered and the mixing fraction of the new quark state is shown as 95\% or more singlet-like (red),  50\%-95\% singlet-like (purple), 50\%-95\% doublet-like (cyan) or 95\% or more doublet-like (blue). In all figures $m_H = m_A = m_{H^\pm} = 1$ TeV. Details of the model, couplings, and decay modes can be found in ref.~\cite{Dermisek:2019vkc}.
}
\label{fig:branching_ratios_doublet_singlet}
\end{minipage}
\end{figure}

In Fig.~\ref{fig:branching_ratios_doublet_singlet}, we show the resulting branching ratios of vector-like top, ($t_{4}$), and bottom ($b_{4}$) quarks to heavy neutral and charged Higgses with $m_H = m_A = m_{H^\pm} = 1$ TeV. $t_{4}$ and $b_{4}$ particles can result from a mixture of doublet- and singlet-like gauge eigenstates. Decays to the CP-odd Higgs are implicitly included. The mixing fraction of the new quark state resulting from the scan of mixing parameters is shown as 95\% or more singlet-like (red),  50\%-95\% singlet-like (purple), 50\%-95\% doublet-like (cyan) or 95\% or more doublet-like (blue). Typical scenarios can easily dominate over decays to $Z, W$ or $h$, where decays to charged Higgs can be close to $100\%$ and for neutral Higgses decays through $H$ and $A$ can each be $50\%$. The impact of these cascade decays for LHC searches is twofold. First, the present limits set by experimental searches through SM boson decays will be relaxed. Second, the signatures of such decays will look very different at current and future colliders, in particular due to the very heavy resonance present in the decay chains.

\section{Signatures at hadron colliders}

\begin{figure}[t]
\begin{minipage}{6in}
\centering
	\raisebox{-0.5\height}{ \includegraphics[scale=0.35]{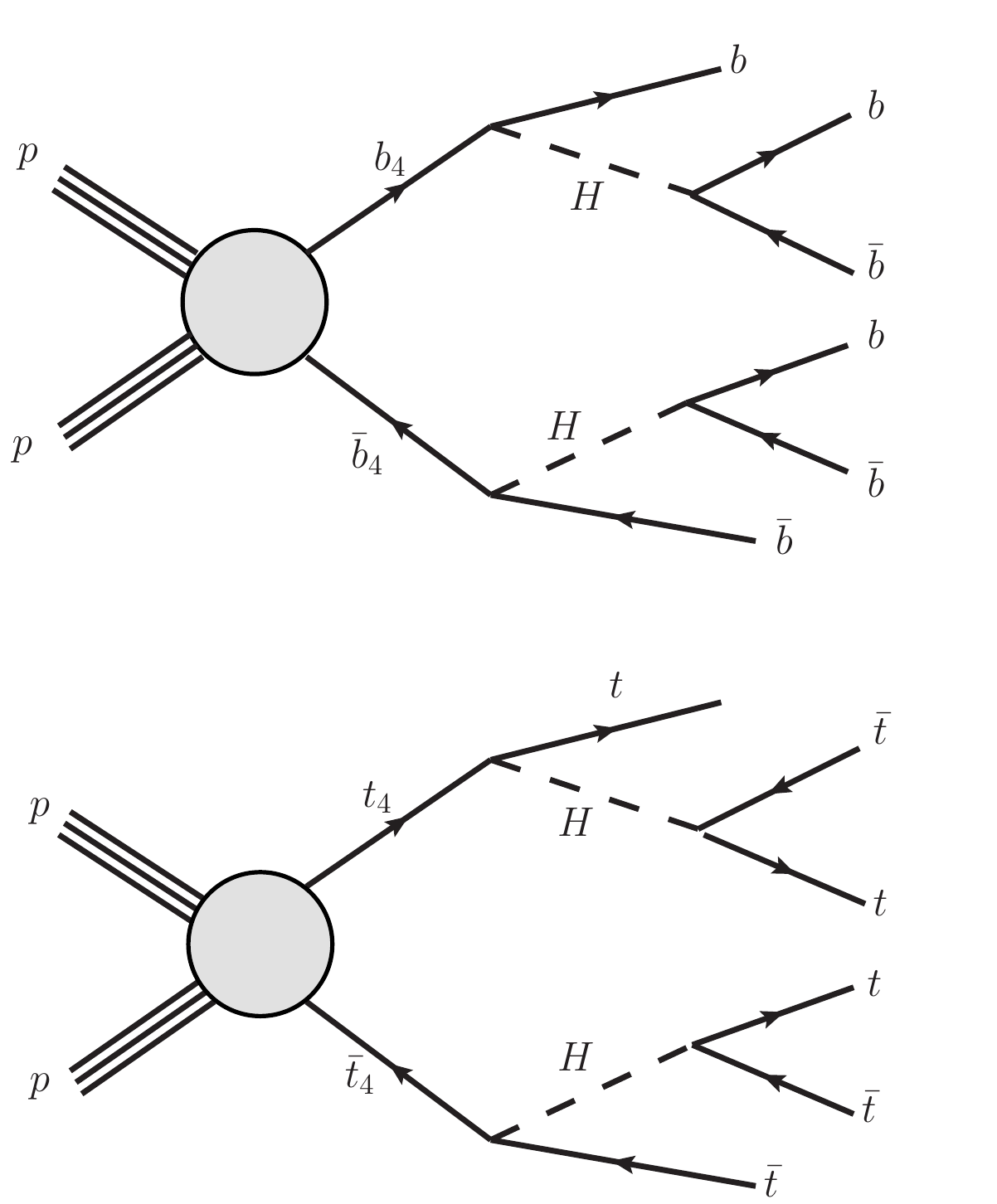} }
  \hspace*{0.1in
  	\raisebox{-0.5\height}{ \includegraphics[scale=0.35]{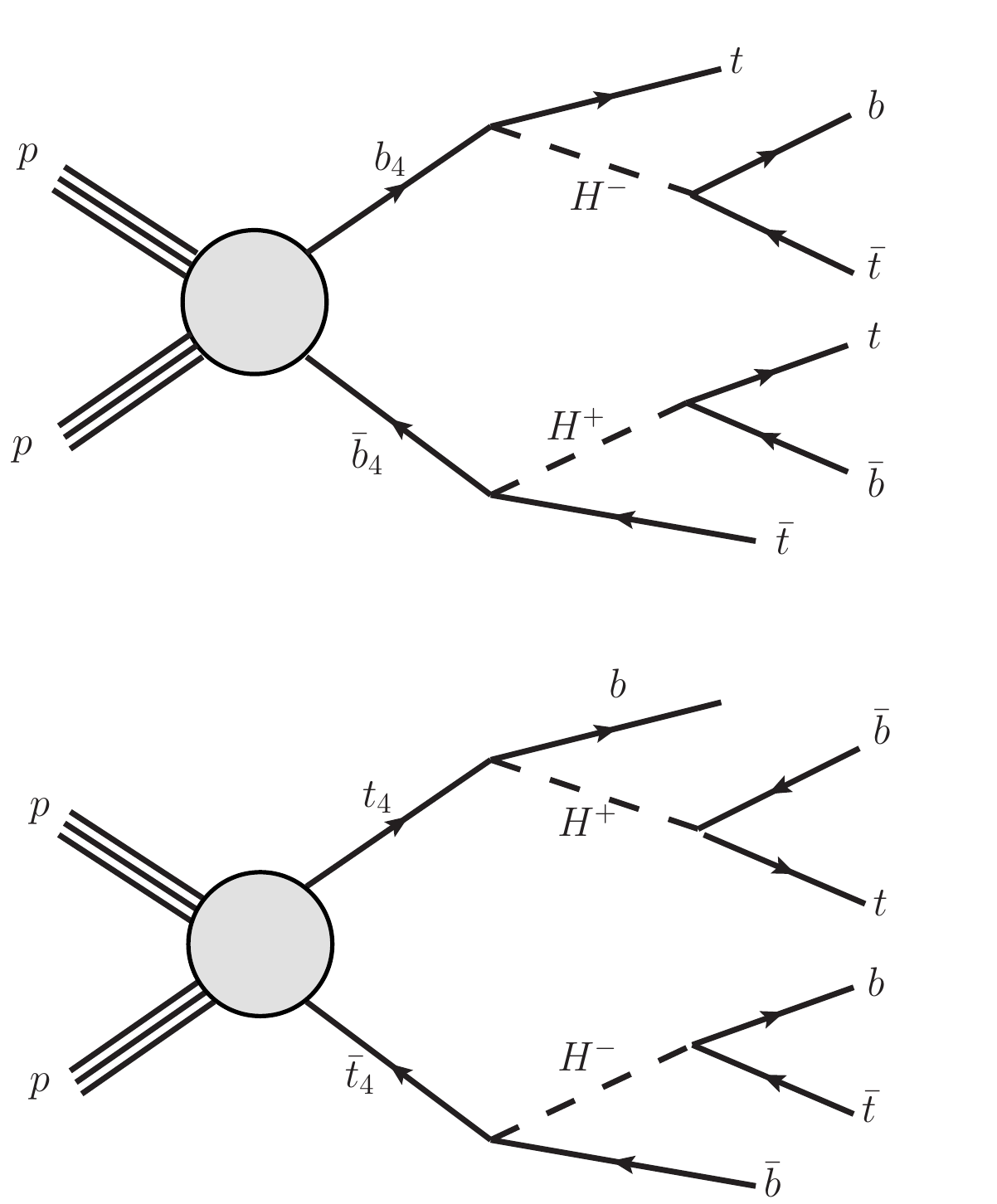} }
    \hspace*{0.1in}
	\raisebox{-0.5\height}{ \includegraphics[scale=0.35]{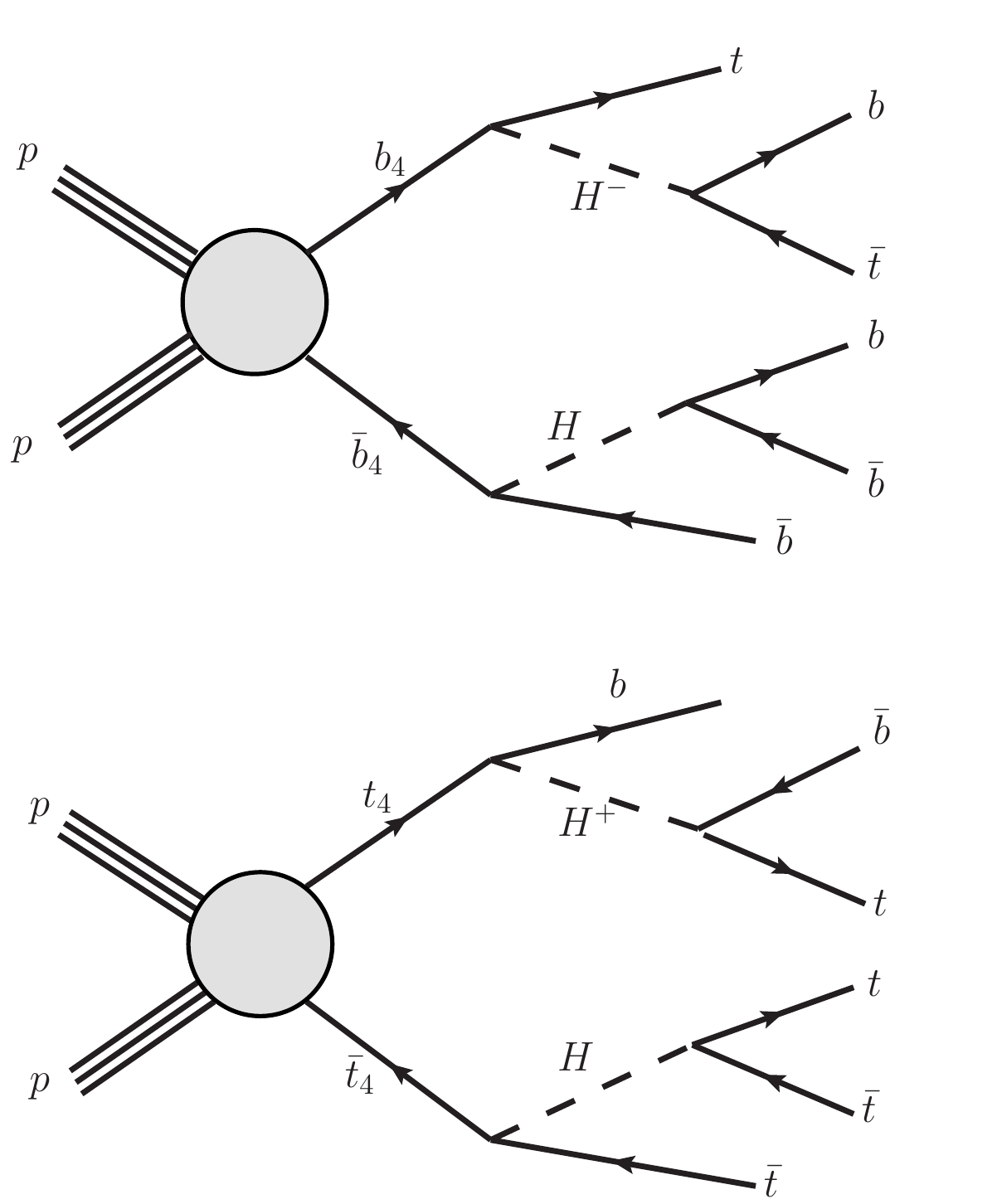}}
\caption{Cascade decays of vector-like quarks through heavy Higgs bosons. Possible channels include decays through neutral Higgses, $H$ (or $A$), charged Higgses $H^{\pm}$, or channels involving charged and neutral Higgses.}
\label{fig:diagrams}
}
\end{minipage}
\end{figure}

In this section, we briefly discuss possible signatures of heavy Higgs cascade decays of vector-like quarks resulting from pair production of vector-like quarks at the LHC. The cross section for pair produced vector-like quarks, generated purely from QCD, depends only on the mass of $t_{4}$ or $b_{4}$. Thus,  for a given vector-like quark mass, one advantage of this channel is that heavy Higgs bosons are effectively pair produced with QCD-size cross sections when branching ratios of vectorlike quarks to heavy Higgses are significant. In Fig.~\ref{fig:diagrams}, we show possible final states at the LHC resulting from vector-like quark cascade decays through heavy Higgses, when heavy Higgses further decay to SM quark pairs. Common final states include multiple top and bottom quarks. Thus, including the dominate decay of the top quark, a common signature to search for vector-like top and bottom quarks, and charged and neutral heavy Higgses is a final state with six bottom quarks.

 Due to the decay of very heavy particles in the decay chains, mulitple, widely separated $b$-jets with significant $p_{T}$, and large global hadronic energies are among the kinematic advantages of the signal over SM background. Possible strategies to extract this signal include analyses based on tagging four or five, high-$p_{T}$ $b$-jets, where the main challenge arises from suppressing irreducible multi-jet backgrounds. For details of the analysis see~\cite{Dermisek:2020gbr}.
 
\begin{figure}[t]
\centering
	\includegraphics[scale=0.55]{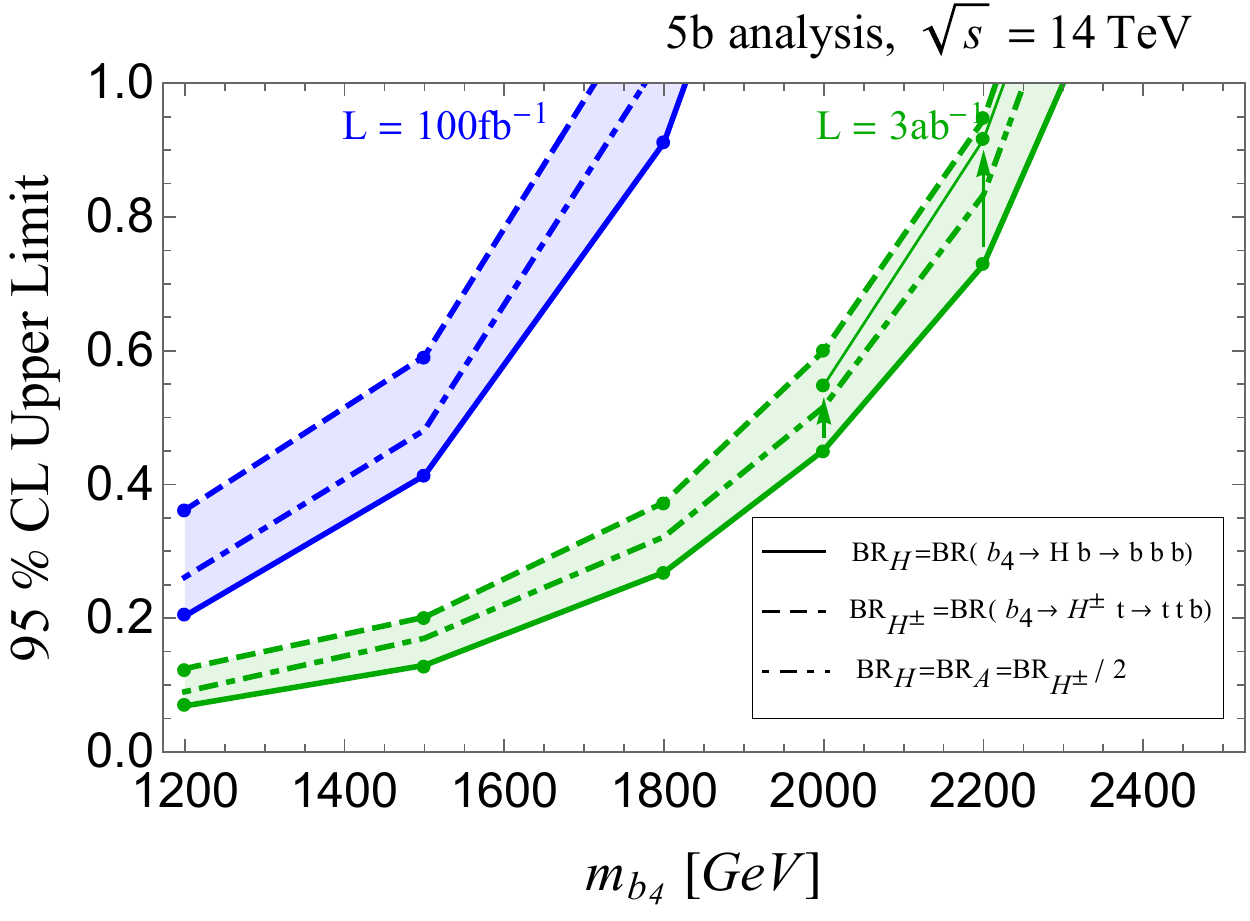} 	\includegraphics[scale=0.55]{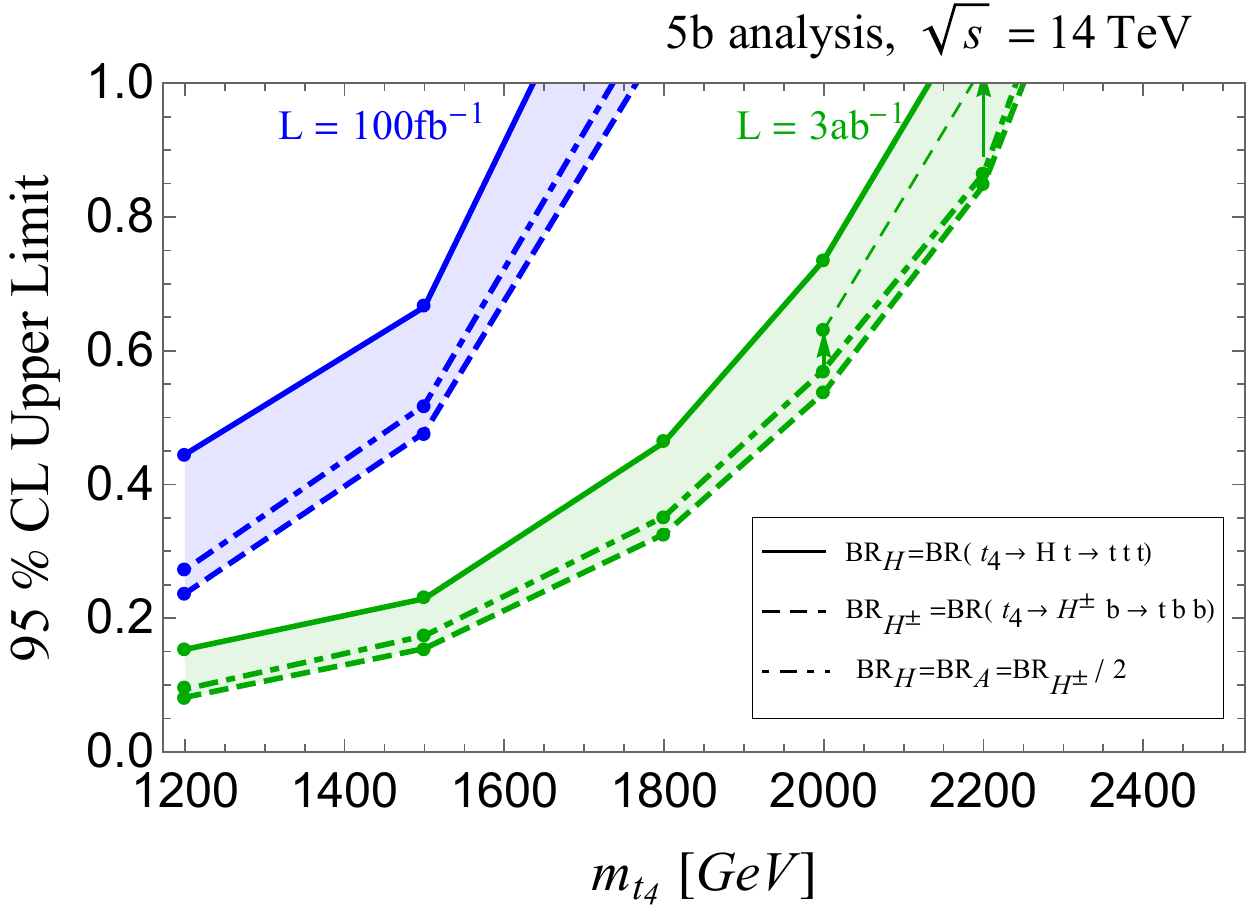} 
\caption{{\bf{Left:}} Projected $95\%$ CL limits on the $b_4$ branching ratio into neutral or charged Higgses for the 14 TeV LHC at luminosities of $L=100~fb^{-1}$ and $L=3~ab^{-1}$. {\bf{Right:}} Projected $95\%$ CL limits on the $t_4$ branching ratio into neutral or charged Higgses for the 14 TeV LHC at luminosities of $L=100~fb^{-1}$ and $L=3~ab^{-1}$. In both figures, solid lines show upper limits on the cascade decay branching ratio for a single neutral Higgs. Dashed lines show the limit for decays involving only charged Higgs. Dot dashed lines show the upper limits in typical scenarios of decays in a type-II two Higgs doublet model.}
\label{fig:5b_b4_t4_bounds}
\end{figure}

 In Fig.~\ref{fig:5b_b4_t4_bounds} we show the projected $95\%$ CL upper limit on branching ratios of vector-like quark cascade decays through heavy Higgses at the 14 TeV HL-LHC in an analysis requiring 5 b-tagged jets in the final state. The solid lines in both panels show upper limits on the branching ratios for a single neutral Higgs. Dashed lines show the limit for decays involving only charged Higgs. Dot dashed lines show the upper limits in typical scenarios of decays in a type-II two Higgs doublet model discussed in the previous section. Cascade decays can probe vector-like quark and heavy Higgs masses up to about 2 TeV for ultimate luminosities at the LHC.

\section{IR fixed points in the MSSM with a vector-like family}
A well-motivated UV completion of the SM which includes an extended Higgs sector and vector-like quarks is the minimal Supersymmetric Standard Model extended with a vector-like family. Recently, in~\cite{Dermisek:2017ihj,Dermisek:2018hxq,Dermisek:2018ujw} several aspects of the renormalization group behavior of gauge and Yukawa couplings in this model have been explored.

\begin{figure}[t]
\centering
	\includegraphics[scale=0.4]{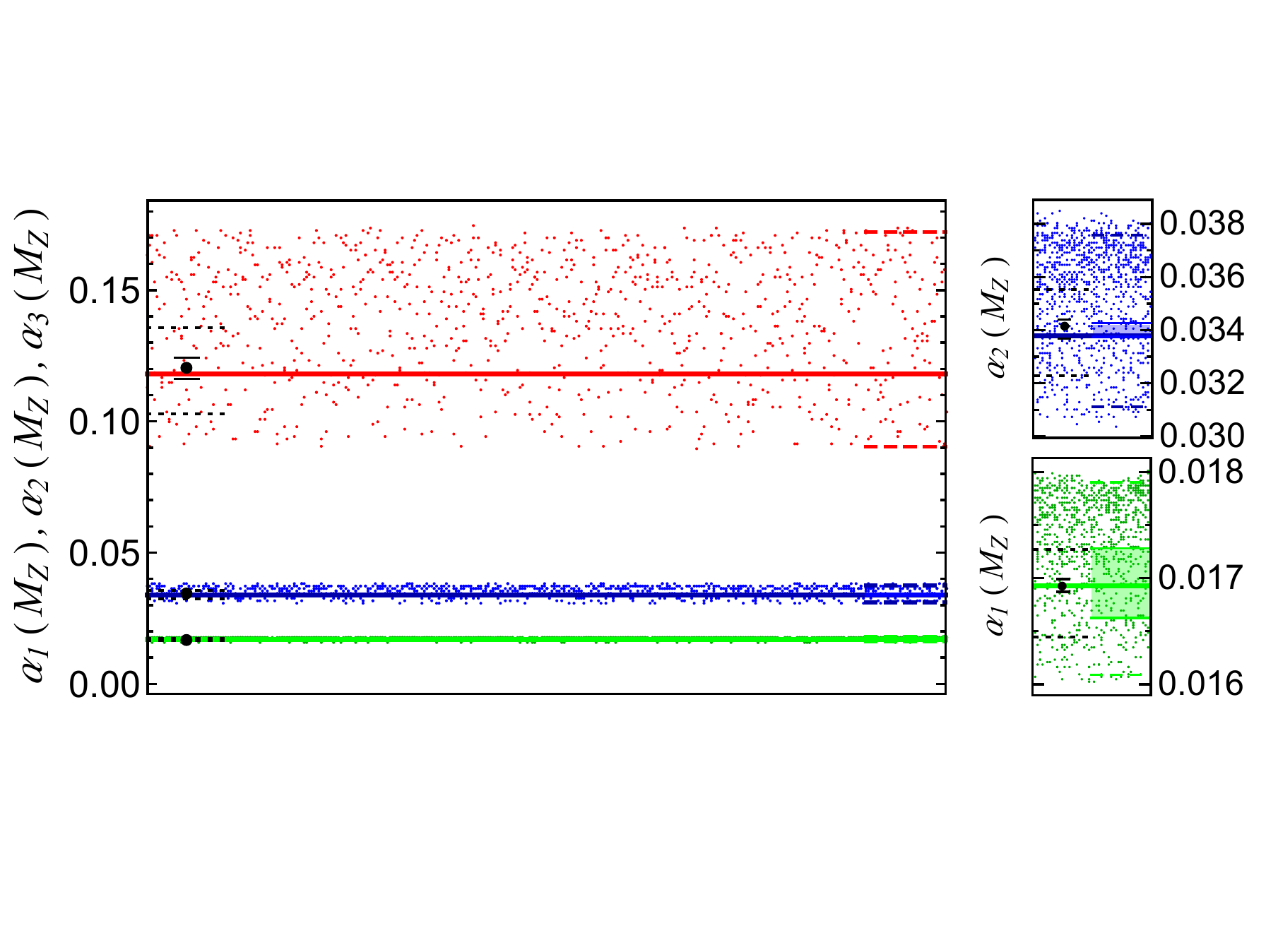} 	\includegraphics[scale=0.4]{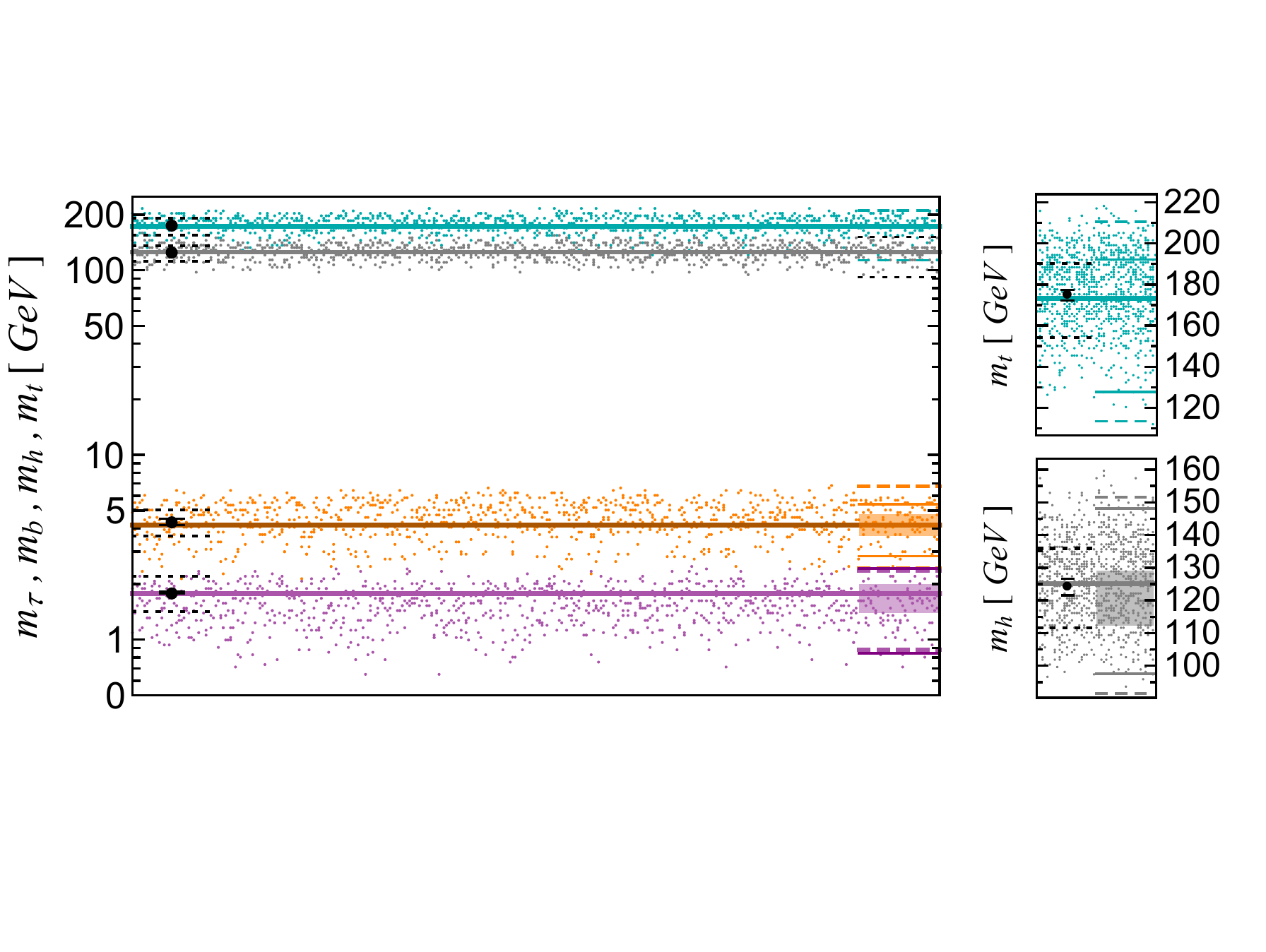} 
\caption{Predicted values of SM gauge couplings at $M_{Z}$, third generation fermion masses, and Higgs boson mass from random scan of couplings in Eq.~\ref{eq:scan}. Solid lines denote the observed central values for all parameters. Dashed lines show the spread of predicted values in GUT scenarios. Thin solid lines show GUT scenarios when $\alpha_{3}$ is fit by varying the universal mass scale $M$. Shaded regions show the spread of predictions when $m_{t}$ is additionally fit using $Y_{V}$. The highlighted points on the left denote the predicted values in the optimized scenario. The spread of predictions from varying $M$ and all dimensionless couplings $\pm 20\%$ is shown in the black solid and dashed lines respectively. }
\label{fig:random}
\end{figure}

The additional matter content in this model drives the RG flow of gauge couplings to large values above the scale of new physics. From the UV perspective, large (but still perturbative) couplings at a fundamental scale, say a GUT scale, flow to trivial IR fixed points. This behavior is inherited in the RG flow of Yukawa couplings. In Fig.\ref{fig:random}, we show the predictions of gauge couplings at $M_{Z}$ and masses of $m_{t},m_{b},m_{\tau},$ and $m_{h}$ from the RG flow of gauge and Yukawa couplings resulting from the random scan of dimensionless parameters:
\begin{flalign}
\label{eq:scan}
\alpha_{1}(M_G), \alpha_{2}(M_G), \alpha_{3}(M_G)&\in [0.1,0.3]\\\nonumber
y_{t},y_{b},y_{\tau} ,Y_{V}&\in [1,3],
\end{flalign}
where $Y_{V}$ represents a universal Yukawa coupling for vector-like fermions. The GUT scale has been taken to be $M_{G}=3.5\times 10^{16}$ GeV, and the scale of all particles beyond the SM is $M=7$ TeV. Solid lines denote the observed central values for all parameters. Dashed lines show the spread of predicted values in GUT scenarios. Thin solid lines show GUT scenarios when $\alpha_{3}$ is fit by varying the universal mass scale $M$. Shaded regions show the spread of predictions when $m_{t}$ is additionally fit using $Y_{V}$. The highlighted points on the left denote the predicted values in one specific scenario, see Fig 3 of~\cite{Dermisek:2018ujw}. The spread of predictions from varying $M$ and all dimensionless couplings $\pm 20\%$ in this case is shown in the black solid and dashed lines respectively. 

We see that completely random values of the seven largest, dimensionless couplings in the SM at the "GUT" scale reproduce the observed pattern of couplings at $M_{Z}$ with only a single scale where all particles beyond the SM content are decoupled. The three fundamental parameters of the model related to mass scales $M_{G}, M$ and $\tan\beta$ can be optimized so that no coupling deviates more than $25\%$ ($15\%$) from its observed values in random (GUT) scenarios. Further optimizing $Y_{V}$ leads to observables within $11\%$ ($7.5\%$) of their measured values. These findings further motivate the search for new particles and dynamics at mulit-TeV scales, in particular scenarios with extended Higgs and matter sectors.

\section{Summary}

Many extensions of the Standard Model include both extended Higgs and matter sectors. In particular, models with vector-like fermions and 2HDMs are often favored for their rich phenomenology and model building possibilities. Experimental searches for vector-like quarks or heavy Higgses typically assume patterns of branching ratios involving only one or the other of these states. However, even small mixing between new physics states and Standard Model particles can lead to drastically different patterns of branching ratios and corresponding signatures at colliders. Final states with six bottom quarks are a distinct signature in models with heavy Higgses and vector-like quarks. This channel can probe cascade decay branching ratios up to masses of 2 TeV for either vector-like quarks or heavy Higgs bosons at the HL-LHC. While we have focused on the interpretation of the analysis in a 2HDM vector-like quarks, the results can also be interpreted more broadly in models involving more exotic bosons, such as a $Z'~\&~W'$, or scenarios of composite Higgs bosons \cite{Han:2018hcu}. Other related studies with vector-like quarks and leptons were presented in~\cite{Dermisek:2015oja,Dermisek:2015hue,Dermisek:2016via,Dermisek:2019heo}

The MSSM extended with a vector-like family provides one such example of TeV scale new physics involving new fermions and an extended Higgs sector.  For mass scales of new particles $\mathcal{O}(10)$ TeV in this model, the seven largest couplings of the Standard Model can be understood for their IR fixed points originating from completely random values at the GUT scale.

Future searches at the LHC will expand our understanding of fundamental physics well into the multi-TeV range of mass scales. Studying combined signatures of cascade decays in extended Higgs and matter sectors has many advantages over traditional search strategies, and opens many new avenues to discovering new particles.

\end{document}